\documentclass[aps,nofootinbib,showpacs]{revtex4-1}
\usepackage[CJKbookmarks, pdftex, bookmarksnumbered, bookmarksopen, colorlinks, citecolor=blue, linkcolor=blue]{hyperref}
\usepackage[english]{babel}
\usepackage{amstext,amsmath,amssymb,amsfonts,bbm}
\usepackage[latin1]{inputenc}
\usepackage{graphicx}
\usepackage{epstopdf}
\usepackage{amsthm}
\usepackage{tocvsec2}
\usepackage{enumerate}
\usepackage{subfigure}
\usepackage{color}
\usepackage{tikz}
\usepackage[squaren,Gray]{SIunits}

 \usepackage{braket}

\newcommand{\be}{\nopagebreak[3]\begin{equation}}
\newcommand{\ee}{\end{equation}}

\newcommand{\bee}{\nopagebreak[3]\begin{equation*}}
\newcommand{\eee}{\end{equation*}}

\newcommand{\ba}{\nopagebreak[3]\begin{eqnarray}}
\newcommand{\ea}{\end{eqnarray}}
\DeclareFontFamily{U}{rsfs}{}         
\DeclareFontShape{U}{rsfs}{m}{n}{<5> rsfs5 <6><7> rsfs7          %
  <8><9><10><10.95><12><14.4><17.28><20.74><24.88> rsfs10}{}     %
\DeclareMathAlphabet{\mathfs}{U}{rsfs}{m}{n}                     %
\newcommand{\n}{{\nonumber}}

\begin{document}

\title{Black holes, Planckian granularity,  and the changing cosmological `constant'}

\author{Alejandro Perez}
\affiliation{{Aix Marseille Univ, Universit\'e de Toulon, CNRS, CPT, Marseille, France}}

\author{Daniel Sudarsky}
\affiliation{Instituto de Ciencias Nucleares, Universidad Nacional Aut\'onoma de M\'exico, M\'exico D.F. 04510, M\'exico}

\date{\today}

\begin{abstract}

In a recent work we have argued that nosy energy momentum diffusion due to space-time discreteness at the Planck scale (naturally expected to arise  from quantum gravity) can be responsible for the generation of a cosmological constant at the electro-weak phase transition era of  the cosmic evolution. Simple dimensional analysis and an effectively Brownian description of the propagation of fundamental particles on  a  granular background yields a cosmological constant  of the order  of  magnitude of the  observed value,  without fine tuning. While the energy diffusion is negligible for matter in standard astrophysical configurations (from  ordinary stars to neutron stars) here we argue that a similar diffusion mechanism could, nonetheless be important for black holes. If such effects are taken into account two observational puzzles might be solved by a single mechanism: the `$H_0$ tension' and the relatively  low rotational spin of the black holes detected via gravitational wave astronomy.

\end{abstract}

\maketitle
\section{Introduction}
  Soon  after  A. Einstein   proposed in  1915  the  general theory of relativity,  it became  clear that its  basic  equations did not allow  for  a space-time  describing a   static  universe, which is   what   was taken as   ``evident" at the times.  
He therefore  introduced  a term  that  changed  that aspect of the theory, and which came to be known as the cosmological constant (CC). The  discovery  in 1929 of Hubble's cosmic expansion seemed to make  such  term unnecessary (although  possible, as long as   its value was  small enough). The development of    the quantum theory of fields   soon  led  researchers to argue that    there  must be a   vacuum  energy  associated with  the zero point fluctuations of all  modes  of all   matter fields,  just  as  the $ \frac12\hbar \omega$   ground state energy  of   an   harmonic  oscillator.  Estimates  indicated  that  the value of  the    resulting  contribution  to the CC    should   be of order $ m_{p}^2$ (where $m_{\rm p}$ denotes the Planck mass). It was  clear very early on that  the   total value of the CC  must be at least  120  orders of magnitude  smaller  than that. A   severe  fine tuning seemed   to  be  required.    Actually,  for a long time  most people  were convinced  that  there was some  deep  physical  principle  that would  explain  why its  value  was in fact vanishing  (see for instance \cite{Hawking:1984hk}).   Things  took a sharp   turn  with  the  discovery  in 1998 \cite{Perlmutter:1998np, Riess:1998cb} of the    fact   that the    universe's  expansion was   accelerating, and   the  realization that  the simplest explanation  was  a rather un-naturally  small,  but non-vanishing   value of the   cosmological   constant. That has in   fact    become  part of   the standard   model  of  cosmology,  also known  as $\Lambda$CDM (involving, besides the well known components representing  ordinary matter and radiation,  a cosmological constant and  cold  dark matter).  The  last couple of  years the   situation  has become  more puzzling   when,  as a result  of an sharp increase  in the precision  of cosmological  observations,   some  inconsistencies  have  appeared    in the value of  the   parameters   characterizing the universe's  expansion extracted from  observations  focusing on different epochs.  One of  the  simplest accounts  for such    observations  would   be  a   small  change in the cosmological constant  from the    era  of   emission of the  cosmic  microwave radiation  about  13.7 billion years  ago, to the  recent  times (corresponding  to the   last few  billion  years).  In the following we  will discuss a  proposal   that   seems to account  in a rather natural way  (i.e. without  fine tuning)   for  both   the  actual value of the cosmological constant,  its recent change,  and  an apparently  disconnected issue:  the unexpectedly low  value    of the  angular momentum of  the black holes that have been observed to  collide   generating  gravity  waves.      

\section{The present  tension   and  our approach}\label{fp}

There is, presently, a $4.4 \sigma$ \cite{Riess:2019cxk} tension between the value of the Hubble expansion parameter today  $H_0$  inferred from the CMB measurements of Planck 2018\cite{Planck 2018}, using  the so-called concordance $\Lambda$CDM model \cite{Ade:2015rim} and the local supernova determination of $H_0$ \cite{Riess:2016jrr}. If this tension were to be confirmed as a failure of the $\Lambda$CDM model, then a natural mechanism for it resolution would consist of having a dark energy component that deviates from an exact  cosmological constant, and which  has grown from the CMB recombination time until today.  Independent  analysis seems to suggest such interpretation \cite{Risaliti:2018reu}.

In a recent series of papers \cite{Josset:2016vrq, Perez:2018wlo, Perez:2017krv} we have proposed a mechanism by which the   usual  ($\Lambda$CDM model)  dark energy component of the  late  universe   can be produced from the noisy diffusion of energy from matter degrees of freedom to the underlying granular structure of spacetime at the Planck scale, taking  place  during the   cosmological radiation  dominated  epoch.    One such  proposal tied the  effect to  an hypotetical   spacetime   granularity which   is   expected from physical properties of black holes in the semiclassical regime\footnote{That is  the   use classical  general relativity coupled with quantum fields in regimes  where quantum field theory on curved spacetimes offers a reliable approximation}\cite{Perez:2017cmj} as well as from various approaches to quantum gravity.  

 On  the  other hand the result in \cite{Collins:2006bw, Collins:2004bp} together   with constraints from the validity of Lorentz symmetry strongly suggest  that any  kind of  Planckian granularity must be of a relational nature, i.e. only apparent when the spacetime geometry is  curved and is  probed with the suitable degrees of freedom, and  thus  its phenomenological manifestation  is severely  restricetd   ( see  for instance \cite{Corichi:2005fw, Bonder:2007bj}). Those considerations  also  suggest   that  the  degrees  of  freedom  that  might be sensitive to Planckian discreteness  should be  massive  field excitations (i.e. ``scale breaking" or ``scale carrying" probes). 
 From  the  macroscale general covariance one also expects that the granularity would manifest itself in a four-dimensionally symmetric manner: {\em `spacetime-grains'} rather than {\em `space-grains'}. It would seem thus  natural   to expect that the probing degrees of freedom  to that  spacetime  granularity,  would be four-dimensional probing excitations,  and that the  effect should  have,  in the   rest frame  associated with  the local matter excitations (or  particles),   a well defined  directionality. Those considerations  led  us to  assume  that  the particles that would  
be  affected  by   such  granular features,  ought,  in addition to  being  massive  also posses  non-vanishing  spin
\footnote{ \label{fn} Scalar massive fields like the inflaton or the Higgs  are one-dimensional probes: they only carry a spacetime arrow given by their four-momenta. Massless (scale invariant) fields are transverse and hence three-dimensional probes: null four-momenta and a spin state in the two-dimensional orthogonal space. Only spinning and massive particles can be seen as genuine four-dimensional probes of the spacetime geometry: their four momenta plus their spin state that can span the  full 4-dimensional `tangent' spacetime. Massless (scale invariant) particles are transverse and hence three-dimensional probes: null four-momenta define their world-line while a spin state probes the 2d orthogonal space. Only spinning and massive particles can be seen as genuine 4d probes of the spacetime geometry: their four momenta tangent to their world-line and their spin state probing the orthogonal 3d space.  }.  Simple dimensional analysis combined with the input that only massive probes might undergo, as well as source, the diffusive effect,  lead,   as described  in \cite{Perez:2017krv}  to an essentially unique law for the deviation from geodesic motion of fundamental particles interacting with the Planckian granular structure
%
The leading term of the force  is:
\be\label{modimodi}
u^{\mu}\nabla_{\mu}  u^{\nu}=\alpha \frac{m}{m^2_{\rm p}}\, {\rm sign}(s\cdot \xi) { \mathbf R}\, {s^{\nu}},  
\ee
 where $\alpha> 0$ is a dimensionless coupling, $m$ is the mass of the particle,  $\mathbf{R}$ is the scalar curvature, $s^\mu$ is the spin of the particle, and $\xi^\mu$ is a preferred time-like vector field characterizing  the  mean state of motion  of the overall matter configuration that is responsible for the spacetime  curvature   (in cosmology this is well approximated by the four velocity of comoving observers). Note that   the  RHS  of  \ref{modimodi}  must be   orthogonal to  $ u^{\nu}$ (from the simple kinematical requirement that $ u_{\nu}u^{\mu}\nabla_{\mu}  u^{\nu} =0$)  and  this  can  easily be  assured   by requiring  it to be proportional to the spin of the particle. This reinforces our  heuristic argument  (see Footnote \ref{fn} and \ref{fn2})  that it should   be  only  the  spin carrying  particles   that might be sensitive to the diffusion mechanism, because the  spin  provides  the only proper  (or canonical)  direction orthogonal to the four-velocity of the probe.

That equation \eqref{modimodi} embodies  `diffusion of energy'  (or an anomalous  violation of  energy conservation)  is clear form the behavior of the mechanical energy $E\equiv -m u^{\nu}\xi_{\nu}$ (defined in the frame  $\xi^{\mu}$) along the particles world-line, namely
\ba\label{conserv}
\dot E&\equiv& -m u^{\mu}\nabla_{\mu}(u^{\nu}\xi_{\nu})\n \\ &=&-\alpha  
\frac{m^2 }{m^2_{\rm p}} |(s\cdot \xi)| {\mathbf R}-m u^{\mu}u^{\nu}\nabla_{(\mu}\xi_{\nu)}.
\ea
The last term on the {\em r.h.s.} of \eqref{conserv} encodes the  standard  change of $E$  associated to the  non-Killing  character of  $\xi^{\mu}$ (e.g., it represents the effect of red-shift in cosmology), while the first term  encodes the friction that damps out any motion with respect to $\xi^\mu$.  Energy is lost due   to the fundamental granularity until $u^{\mu}=\xi^{\mu}$ and the particle is at rest with the cosmological fluid, and thus the  anomalous  effect would cease.  \footnote{It is unclear if one  should think of this energy  as   being `transferred'  to quantum gravity degrees of freedom, or rather lost in entities without any permanence  (in time)  as would be the case for spacetime `defects'.  On the other hand,  it is  worth noting that   general considerations discussed in \cite{Maudlin:2019bje}  indicate  that  energy conservation is a  notion which,  strictly speaking  (i.e.  as an  exact statement),  we  would  probably be forced to forgo. However, if those entities correspond to degrees of freedom in usual sense of the word (with permanence in time) they could be relevant in dealing with the information loss puzzle in black hole evaporation \cite{Perez:2014xca, Amadei:2019ssp}.}.  Note however that,  in principle  one might have $ \xi.s =0$ even  when  $u$ is not parallel to  $ \xi$,   but 
as far as a gas  of particles  in thermal equilibrium is concerned ( i.e. the situation consider in the  previous  works)  that   condition applies  only to a set of  measure  zero and in fact  the  inter-particle collisions  will   rapidly   take  any particle away  from that  set.

At this point we  should note that   self-consistency of the proposal requires  that, together with
 equation  \ref{modimodi} above,   one  introduces  a modification for  the evolution    equation  for the spin,   simply because by construction  we ought to have  $u\cdot s=0$ at all times. 
In fact  the  simplest   expression at the  same order of analysis as  that  of  equation  \ref{modimodi}
is  given  by:
\be\label{spinny} u^{\mu}\nabla_{\mu} s^{\nu}=\alpha \frac{m}{m^2_{\rm p}}\, {\rm sign}(s\cdot \xi) { \mathbf R}\, (s\cdot s)\, u^{\nu} .\ee
This is just a minimalistic solution, other terms can be added if one  makes  further use of  the direction $\xi^a$.

 These ideas  were  used  in a previous  work \cite{ Perez:2018wlo, Perez:2017krv}  to   provide  an account  for the    nature  and magnitude of the  dark energy  component  that  became  dominant in the latter  epochs of  the cosmic  evolution. The point is that although  standard  general relativity   is  inconsistent  with  any violation of the  conservation of     energy momentum ($\nabla_{\mu} {\mathbf T}^{\mu \nu} =0$),  a    relatively mild  modification known as  unimodular  relativity    can  readily accommodate it (provided  a certain  integrability condition holds). The equations of  uni-modular gravity are:
    \be 
\mathbf R_{ab}  -\frac14  g_{ab} \mathbf  R  =  8\pi G   \left(\mathbf T_{ab}  -\frac14 g_{ab} \mathbf  T \right) 
\ee
   which   as we noted  admits ${\mathbf J}_a\equiv  8\pi G \nabla^b \mathbf  T_{ab}  \not=0$.    After some  simple  manipulations  (and making use  of the  standard  Bianchi  identity), and assuming that $dJ=0$,  one finds, 
\be\label{Lambda-Eff}
\mathbf R_{ab} -  \frac12 g_{ab}\mathbf  R   +   g_{ab}   
\left(   \lambda_{0}  +  \int {\mathbf J}   \right)
 =  8\pi G   {\mathbf  T_{ab}} 
\ee
 with    $\lambda_{0}$  an integration  constant.  
 In  the  homogeneous and isotropic cosmological setting the 
 integrability conditions hold automatically.  The  point is of course that  $ \lambda_{0}  +  \int {\mathbf J} $  will act,  at late times,  when  $\mathbf J$  becomes  negligible, as  an effective cosmological constant. Simple use of kinetic theory allows  for the  evaluation of  $\mathbf J$  resulting from our  postulated  friction  effect  during    the relevant cosmological   epochs  where the effect is  strongest. This procedure,  together with  the assumption that a  protective symmetry  operating  at  the  Planck  epoch  fixes   $\lambda_{0}=0$,   leads to an  estimate of  the  effective cosmological  constant whose   order of magnitude  naturally (i.e.  with  $  \alpha\sim O(1)$  and  no fine-tuning)  matches the observations.   The detailed  study reveals that the  effect  is overwhelmingly dominated by the contributions  coming from the epoch  when particles first  acquire mass, namely the electro-weak  transition.

We  ought to point out that the  term  ${\rm sign}(s\cdot \xi) $  in   equation \eqref{spinny} has  a non differentiable behaviour in the 
 limit where $u$ becomes colinear with $\xi$ and $s\cdot \xi\to 0$. This corresponds to the (non-relativistic) limit where the kinetic energy of the particle, as measured by co-moving observers, goes to zero. The form of this factor in this limit was not important in the analysis of \cite{Perez:2018wlo, Perez:2017krv} because    in that case  the contributions to  the formula were always dominated by  effects  coming  from  the relativistic regimes  (temperatures higher than,  or of the order of  the mass of the particles involved).  In  effect it is  natural  to expect  that quantum field theoretical modeling of the effect  
would  lead to  modified  versions  with  a smooth  low energy   behavior, which would  however result in   similar effects in   settings involving  relativistic  particles.  An example  is provided  by the replacement of such term by   $(s\cdot \xi/u\cdot \xi)$  which behaves as  ${\rm sign}(s\cdot \xi) $   for  $ u^{\mu}$  corresponding  to ultra-relativistic motion  as seen in the frame defined by $\xi^{\mu}$. Other natural regularizations  are  provided by expressions such as    $(s\cdot \xi/u\cdot \xi )^{2n+1}$ for a natural number $n$.  None of those   variants   would  modify the results  as  far  as the generation  of an effective cosmological constant   during the  electro-weak transition era  in  cosmology.   As  we  will see, however,  those  modifications could have important impact in  the   low  energy regimes, and  in particular in the manifestation of  the type of effect  in question  during the latter stages  in the  universe's evolution.

\subsection{Macroscopic objects and the Planckian diffusion force}

 We now  consider possible effects that the force implied by equation  \eqref{modimodi}  could produce on macroscopic objects.
The first issue is  what  is  the appropriate  manner  in which such  question must be addressed.  The point is that in the context of GR the motion of extended  objects is, in principle, a highly complex  matter  requiring for instance a suitable notion of macrosocopic localization (e.g. center of mass) and  the   study of  the manner in which it evolves\cite{Bigelbock}. The point is that even in highly idealized situations  the motion  of  something like a center of mass  in GR  does not correspond to  a geodesic world-line,  but  is  instead  described  (to the lowest order  in curvature corrections) by effective equations like, for example, the
Papapetrou  equation \cite{Papapetrou:1951pa}.  The general  treatment of these  issues,   is thus  highly non-trivial, as   follows from the fact that no  simple  covariant  notion of `addition of  forces'   to compute the total force, nor `addition of positions  weighted  by mass' to compute a suitable `center of mass' position\footnote{Considerations   motivated  by the  fact that in  the real world,  the   characterization  of a  particle as truly a  point-like  object  (with  infinitely  precise localization  at  all times), cannot  be taken as anything   more than a  very  good  approximation, lead  to  rather interesting  conclusion as  about  the   nature of  geometry as an emergent   entity \cite{Bonder:2017jcu}.}.   In the regimes where the  weak  field   approximations  hold,  the question  simplifies  dramatically, and that  is in fact  what  we  encounter  in most practical situations  when dealing with GR.
  
Let  us for instance consider  two  point  particles  undergoing  the effect   described  by equation  (\ref{modimodi}) in a  situation where  the spacetime  geometry is well approximated by Minkowski and the relative velocities of the particles are small so that a Newtonian analysis is justified.  The force on the center of mass is controlled by the sum of the forces,  but that is  very different  from the  force  one  would estimate  by considering  the whole system  as a point particle placed at the center of mass undergoing  the effect of equation (\ref{modimodi}). This follows,  among other reasons, just  from the fact  that  the angular momentum of  the  composite object  about  its  center of mass is  not  simply the sum of  the spins of the constituent particles. 
 It is therefore  clear that,  even  if  equation (\ref{modimodi})   applies to the  single elementary particles,  it cannot be taken   as a general recipe for the  deviation from  geodesic  motion for an extended object. In that case  the best  manageable estimate of the effect  is obtained  by applying   equation (\ref{modimodi})  to each   individual constituent and then combining the effects in the appropriate manner, be it in a  Newtonian or  in a fully relativistic regime (where an  analysis analogue to that of  Papapetrou's\cite{Papapetrou:1951pa} would have to be  used).
 
Let us consider  next the  magnitude  of  such effects for some  simple astrophysical objects. As we noted in the `Newtonian regime'  we should simply add the  forces  on all the constituents of the macroscopic body.  As the force is proportional to the scalar curvature (that is dominated by the density for non-relativistic matter) it seems that the ideal candidate for macroscopic astrophysical object maximizing our force would be neutron stars (or quark stars). Let us estimate the force in that case.
We will also assume that the neutrons are highly polarized along a common direction. This is the configuration where the individual particle contributions would contribute coherently. It is certainly not representative of the typical astrophysical situation but it gives an upper bound for our force.
It follows (using   $|s| =1/2$) that
\be
|F|\le \alpha  \sum_{i}\frac{m_i^2}{m_{\rm p}^2} {\mathbf R}=\alpha \int n \frac{m^2}{m_{\rm p}^2} {\mathbf R} dV,
\ee
where in the first equation the sum is over constituent particles (say neutrons) with masses $m_i$, and in the second line we introduce the number density $n$ of particles and assumed that $m_i=m$ for all particles.  The inequality follows in part from the statistical fluctuations in the behavior of what we wrote as ${\rm sigm}(s\cdot \xi)$ in \eqref{forza}, and the fact that in realistic  situations  not all the individual forces will be aligned. As noted above   in the low  energy  regime such simple characterization  of the anomalous    force requires some regularization near $s\cdot \xi=0$ which would  naturally  make  the total  force  even smaller. Ignoring for the moment all the above caveats,  using that $\rho=m n$ and assuming the $\mathbf R$ and $\rho$ are constants we can now write
\be
|F|\le \alpha \frac{m M}{m_{\rm p}^2} {\mathbf R} 
= \alpha\frac{m M}{m_{\rm p}^2} {\frac{8\pi \rho}{m_{\rm p}^2}}
= \alpha8\pi \frac{m M}{m_{\rm p}^2} {\frac{m^4}{m_{\rm p}^4}} m_{\rm p}^2,
\ee
where $M=\int \rho dV$ is the total mass of the star. In the second line we used Einstein's equations, and in the last line we used that $\rho\approx 1 ({\rm GeV})^4\approx m^{4}$ as would  be appropriate for a neutron star. In order to 
characterize  the   relevance  of the   effect  through a  dimensionless quantity, we can compare  the  new force  with  the  usual  `gravitational  force' binding   the neutron star to  a galaxy like ours.  Such comparison provides an indication of  whether the Planckian friction can generate  substantial deviations  from the motion of the neutron star away from its expected Newtonian trajectory. For a galaxy with  mass $M_{\rm g}$, and  using  the {\it virial}  theorem, we know  that (in average) we  have,
\be
M v^2\approx \frac{G M M_{\rm g}}{R},
\ee
where $v$ is the (mean) magnitude of the velocity of the neutron star. From the previous equation  we find  that $R\approx GM_{\rm g}/v^2$
The gravitational force is then
\be\label{galaxy}
|F_{\rm g}|=\frac{G M M_{\rm g}}{R^2}
=\frac{M v^4}{G M_{\rm g}}
=\frac{M v^4}{M_{\rm g}} m_{\rm p}^2.
\ee
We can now compute the quantity of interest, namely
\be
\left|\frac{F}{F_{\rm g}}\right| \le  \alpha8\pi \frac{m M}{m_{\rm p}^2} {\frac{m^4}{m_{\rm p}^4}}\frac{M_{\rm g}}{M v^4}
= \alpha8\pi v^{-4} \frac{m^5}{m^5_{\rm p}}\frac{M_{\rm g}}{m_{\rm p}} 
\approx \alpha  (2 \times 10)(10^{12})(10^{-5\times 19}) (10^{12} 10^{38}),
\ee
where we used  $v\approx 10^{-3}$, $M_{\rm g}\approx 10^{12} M_{\odot}$ and  $M_{\odot}\approx10^{38} m_{\rm p}$. The result is then
\be
\boxed{\left|\frac{F}{F_{\rm g}}\right|\le 2 \alpha \times 10^{-32}}
\ee  
Thus  for $ \alpha \sim O (1) $ the force is negligibly small in comparison  with the gravitational binding force for a neutron star in a galaxy. It seems clear that the previous bound will apply to any other star or macroscopic object for which the density is bounded by that  of the neutron star,  and for which spins will tent to average out  to  zero, leading to  a  decrease  in the   value of the   total   cumulative  force  resulting form the  ones  described  by  equation   \eqref{modimodi}   acting on  each   elementary particle  constituent.


\section{Granularity and diffusive effects in black holes}

The successful  prediction of the value of the  Dark Energy component  described in \cite{Perez:2018wlo, 
Perez:2017krv},  corresponding  to    the generation of  an effective cosmological constant    during the  electro-weak  epoch,   and   which remains   essentially constant afterwards,   naturally suggest   the  search for a similar account  for   the   recent observations that   indicate  that  the dark energy  content might have  changed from   recombination time  until today. However,  as the universe has cooled down about $17$ orders of magnitude from the EW transition epoch the effects of the diffusion (\ref{modimodi}) applied to fundamental particles has become completely negligible.  Thus that mechanism cannot produce the type of late time growth of $\Lambda$ that is indicated by the observations that lead to the  $H_0$ tension. Nonetheless, a new (possibly unprecedented) situation of high density has become relevant in the later evolution of the universe with the appearance of galaxies and stars: this correspond to the occurrence of gravitational collapse leading to BH formation\footnote{Although  speculative ideas   about the possibility of  primordial  black holes  forming in  much  earlier times  have certainly been contemplated.}.


\subsection{Black holes and translational energy  diffusion  from  relational  space-time  granularity}

Black holes share many features with fundamental particles in their relation with  the  external  environment: when isolated they are completely characterized by their mass, charge and spin with a gyromagnetic factor $g$ with the same value as the one predicted by the Dirac equation \cite{PhysRev.174.1559} \footnote{On the other hand the  considerations  that   allow  the   introduction of   CoM  in   general relativitistic   contexts dealing with  extended  mater distributions\cite{Bigelbock, Papapetrou:1951pa} 
 are grossly violated in the case of  Black Holes, as  clear-cut evidence of that we  can point to the fact that such objects can exist and be produced as vacuum solutions.}.
This  suggest  that black holes  might  be the only  kind of macroscopic  objects for which the effects of the space-time granularity explored here could be described by an equation like (\ref{modimodi}) or something  very close to it.  Thus  we will   again  postulate  that, as a result of the space-time  granularity  of  quantum gravitational   origin, black holes  would be affected by  a  friction-like  term  with the form:   
\be\label{modimodi2}
u^{\mu}\nabla_{\mu}  u^{\nu}= \overline\alpha_{\rm bh} \frac{M}{m^2_{\rm p}}\, {\rm sign}(s\cdot \xi) \tilde { \mathbf R}\, {s^{\nu}},  
\ee
where the new phenomenological dimensionless coupling $\overline\alpha_{\rm bh}$  has been introduced. The point is  that  as a result of a possible Wilsonian `running', or averaging procedure due to the macroscopic character of the black hole---keeping in mind the emergent nature of spacetime geometry advocated here--- that parameter need not be the same as the $\alpha$ for fundamental particles in equation \eqref{modimodi}.  
That the new coupling $ \overline\alpha_{\rm bh} $ need not be the same seems clear from the fact that its derivation would not only involve a fundamental quantum gravity theory  describing the    relational discreteness, but also  the  detailed  structure of  macroscopic black holes in that theory. The  factor $\tilde { \mathbf R}$ in the present case  must be regarded  as  some average  measure of the space-time  curvature  associated  with the  matter specifying the locally-preferential  frame   in the  surroundings   of  the black hole\footnote{This  is  a delicate and complex  conceptual  issue,   and  we  do not  claim to have a clear control of  all its ramifications  as of  yet.   The  point is that while it is true that  GR  is  a purely local theory,   we  can expect  quantum gravity to   involve   both,   some of the  local aspects of GR,   but also  some  of the non-locality inherent to quantum theory, as demonstrated by  the violation of Bell's  inequalities.  Thus  the relational aspects of the   quantum gravity granular characteristics   of space-time  might  be  expected to  involve  aspects  that  reflect   a   certain degree on non-locality.   Of course,   just as in the case   of  EPR  and the situations  considered  in Bell's results,   we   would  expect those nonlocal  aspects   to   coexist  peacefully  with relativity in the sense   of   not allowing  faster than light communication. We will not  pursue  that  interesting   discussion further here  but can point the intersted reader  to  works such as \cite{Myrvold, Bedingham:2007zz, Pearle:2014lha, Tumulka2006}  for more analysis.},  that is,  it must  be intimately  tied to $\xi^\mu$ .

 We   will next  consider  the  effects of  such  proposal  for   various  processes    occurring in the universe  as a result of  the late time formation of  structure, and  eventually,  on  value of the effective cosmological constant. That is  we  will consider   black holes as  the analogues  of
 fundamental objects   undergoing a similar effect   at late times,   as  the elementary particles did  in the  electro-weak transition epoch,   in the   proposal  contemplated in \cite{Perez:2018wlo, 
Perez:2017krv}.
However before  any further analysis of that idea,   we  should  obtain a rough estimate   of the  astrophysical  importance  of the effect of  equation (\ref{modimodi2}),  by comparing its   magnitude  with that resulting  from the  relevant  `` gravitational ( Newtonian)  forces"  in two interesting scenarios:

 First we compare its order of magnitude with   that of  the gravitational force at the moment of closest approach in black hole coalescence, $F_{\rm BHC}$. We estimate $F_{\rm BHC}$ for two equal mass BHs with the Newtonian expression when the separation is of the order of the Schwarzschild radius and the black hoes  are extremal so  $s= (M/m_p)^2$. This gives $|F_{\rm BHC}|\approx G M^2/(2GM)^2=m_{\rm p}^2/4$ from which it follows that
\be\label{GW}
\left|\frac{F}{F_{\rm BHC}}\right|\le 4\overline\alpha_{\rm bh} \left(\frac{M}{m_{\rm p}}\right)^4 \frac{\tilde{\mathbf{R}}}{m_{\rm p}^2} =
4\overline\alpha_{\rm bh} 10^{32} \left(\frac{M}{M_{\odot}}\right)^4,
\ee  
where we have used  $\mathbf R\approx 10^{-120} m_{\rm p}^2$, i.e., we have assumed that the mean scalar curvature around the black hole comes entirely from the contribution of the cosmological constant; we are assuming for concreteness that the environment of the black hole is not contributing to the definition of the curvature entering \eqref{modimodi2}\footnote{\label{dragging-frames} It  is   of  course   possible  to consider  our  kind of model  in which  it is  the galactic  matter  that  is  providing  the source of ${\mathbf{R}}$ in  equation  (\ref{GW}), but in  that case,  consistency  with the relational view  described in  \cite{Perez:2018wlo, 
Perez:2017krv}, would   require that the  local preferential frame  $\xi^\mu$ associated  with the granular  space-time structure,   would also be  determined by the `mean notion' of the local  matter  distribution.  In  the case  at hand that matter   is  that which  is  presumably   differentially  rotating with the galaxy,  thus   making the  exploration of the  issue   a much more complex task.  For instance,   one would have to  consider  the various possibilities  for what exactly one is  supposed to  take  as the `mean motion'   mentioned   above    and/ or   the   size of the region one is supposed to   average over in  determining that.  The point is,   that in  such  case,   it   is   rather unclear   what    would   it  mean  exactly,  for a black hole to  be `slowed  down'  by the   granularity induced friction,  in  its motion   relative to the  surrounding   gravitating media.  That kind of  study,  although  in our view   rather worthwhile,  falls  well outside the scope  of what we intend to to do in the present   manuscript.} . 
The point is  that this   already gives an extremely  large effect if $\overline\alpha_{\rm bh}$ would be order one,  which  strongly   suggest   that if the effect is there   at all  the dimensionless coupling $\overline\alpha_{\rm bh}$ must be very small. We note however   that the above   estimation concerns  extremal black holes  and  it would  be negligible for BHs with vanishing rotation (see  also the discussion  in   the   next  section). Moreover  we  note,   that as   discussed in  the Footnote \ref{dragging-frames},  it would not be  surprising if  the    general  relativistic  `dragging of the frames'  was  somehow   also   affecting the granular space-time structure and  thus  came to play a role in  this  situation.  In the local frame of `quasilocal observers' near the horizon the black hole is seen as  non-rotating (this is apparent in the analysis of \cite{Frodden:2011eb}). That could play a role in accounting for the suppression mechanism for the force as described in macroscopic terms.     
 Again ignoring such  caveats, we will explore two natural possibilities for suppression in what follows: the first one where we assume that 
\be\label{GW1} \boxed{\overline\alpha_{\rm bh}=\overline\alpha_{\rm bh}^{01} {\frac{m_{\rm p}}{M}}\ \ \ \Longrightarrow \ \ \ \left|\frac{F}{F_{\rm BHC}}\right|\le
4\overline\alpha^{01}_{\rm bh} 10^{-6} \left(\frac{M}{M_{\odot}}\right)^3},\ee
and a less suppressed possibility where
\be\label{GW2}\boxed{ \overline\alpha_{\rm bh}=\overline\alpha_{\rm bh}^{02} \sqrt{\frac{m_{\rm p}}{M}}\ \  \Longrightarrow \ \ \left|\frac{F}{F_{\rm BHC}}\right|\le
4\overline\alpha^{02}_{\rm bh} 10^{13} \left(\frac{M}{M_{\odot}}\right)^{\frac72}}.\ee
Note that this type of suppression  could arise naturally from some fundamental description if the process is mediated by some stochastic process where the number of `area quanta' $N_A\equiv M^2/m^2_{\rm p}$ or the  number of `energy-quanta' $N_M\equiv M/m_{\rm p}$ plays a role. In such case the supression we are considering in $\overline\alpha_{\rm bh}$ would be the usual stochastic factor $1/\sqrt{N}$. Such deviations from classical expressions arises naturally in some studies of BH entropy calculations in loop quantum gravity \cite{Ghosh:2013iwa}. Commenting on the immediate apparent phenomenological implications,   while the second possibility seems  more susceptible to  being  to ruled out empirically, it is not obviously that it is  at this point  in contradiction with observations if the mergers observed so far have corresponded to non spinning black holes.  However, in the  second  case  and    unless   $a/M<10^{-13}$   a   strong   suppression in  $ \overline\alpha_{\rm bh}^{02}$   would  be  implied  by the fact that  data  from  LIGO  has   so  far come  out  very    well    adjusted to the   standard expectations from pure GR. 

We can also compare the force \eqref{modimodi}  to the Newtonian gravitational force binding the BH to its galaxy \eqref{galaxy}. Then we get the previous estimate enhanced by the factor ${M_{\rm g}}/ (4 {M v^4})$, and the non relativistic factor $v^{2n+1}$ (recall discussion at the end of Section \ref{fp}) namely
\be
\left|\frac{F}{F_{g}}\right|\le \overline\alpha_{\rm bh} 10^{44}\left(\frac{M}{M_{\odot}}\right)^3 v^{2n-3}= \overline\alpha_{\rm bh} 10^{53-6n} \left(\frac{M}{M_{\odot}}\right)^3,\ee
where we used $M_{\rm g}\approx 10^{12} M_{\odot}$, $M_{\odot}\approx10^{38} m_{\rm p}$, and we  used $v\approx 10^{-3}$. 
 Thus,  once more,   unless   $\overline\alpha_{\rm bh}<  10^{-56+8n}$ 
 the model  would  imply large deviations  from  the standard predictions  of  general relativity for  spinning BHs of intermediate masses.  
In the two possible scenarios proposed in \eqref{GW1} and \eqref{GW2} we get
\be\label{gala}
\boxed{\left|\frac{F}{F_{g}}\right|\le \overline\alpha_{\rm bh}^{01} 10^{15-6n} \left(\frac{M}{M_{\odot}}\right)^2},\ee
and
\be
\boxed{\left|\frac{F}{F_{g}}\right|\le \overline\alpha_{\rm bh}^{02} 10^{34-6n} \left(\frac{M}{M_{\odot}}\right)^\frac{5}{2}}.\ee
Note that the effect of the integer $n$ controlling the low energy smoothing of the ${\rm sign}$ function in \eqref{modimodi} has an important effect in the order of magnitude of the effects, and can, for low values---like $n=3$ in the first case, and $n=6$ in the second---completely dominate over the other large factors.

We note however  that all the  above  bounds have been computed with the assumption that the BH is close to maximally rotating, i.e. the assumption that
\be
s_{\rm BH}\approx  \left(\frac{M}{m_{\rm p}}\right)^2=10^{76}  \left(\frac{M}{M_\odot}\right)^2. 
\ee 
The force is enhanced by this huge factor for highly rotating BHs but it could be   substantially   reduced  for  black  holes that   have  negligible rotation. We  will   consider  below a  mechanism that would  in fact  all  but assure that this  this  would be  the case for essentially of all  black holes  we  would   find,   with   the  only  exception being those   black holes  that  have  just   been involved in  a  collision.   This   means that  a  very interesting   path  for     the   study of  our  proposal, and    setting   important constraint on the model's parameters,   is open  by  detailed studies of the  post collision  part  of the gravity wave  signals,  including the ``ring down", which  is  allready  leading to constarints on deviations from   GR at the  10\%  level \cite{Isi:2019aib}.  

We will discuss the possible phenomenological implications of this effect in galaxies in Section \ref{peno}.
When it comes to the contribution of such forces to cosmological evolution what one needs to estimate is the total amount of energy that the mechanism can actually diffuse,  and thus,  according to the  general  ideas    developed in  
\cite{Josset:2016vrq, Perez:2018wlo, Perez:2017krv},  bring  about  a  change  in  the effective value of cosmological constant. It is clear that  a precise analysis would necessitate a  solid  understanding of the cosmological  distributions of   black holes  as  a  function  of  masses,   spins   and  velocities  and their  evolution throughout cosmic  history.   
 
However,  it is  easy to see that  the present  effect  would  naturally be overshadowed  by  another   one, namely the one associated  with  rotational friction we  will discuss  next. The argument is  rather simple,  independently of the value of the parameter  $\overline\alpha_{\rm bh}$   the  effect   on the cosmological constant 
  is  bound  by the magnitude  of   kinetic   energy   associated  with the peculiar  velocities of  the  black  holes  relative to  the frame  defined  by their local gravitational environments   (the mean motion of all matter in their  environment).   The black holes  would  be  generically moving with non-relativistic  velocities 
  of the order  of  the mean velocities of stars in galaxies, i.e.  $v\sim 10^2  {\rm km \, s^{-1}} \sim 10 ^{-3} c$. Therefore  each   each black hole   
  could contribute at  most  with     something of the order of  $ \frac12 M_{\rm BH}  v^2  \sim    M_{\rm BH}  10^{-6}  $.  
  That is,   there  would  be  at most 
  $  10^{-6}$  of $\rho_{\rm BH}$,  the  energy density in black holes,  as  an available  source to contribute to  our   effect.   On the other hand a  rapidly  rotating  black hole  can  
   in principle  lose a  substantial  fraction of  its total energy   if  a friction-like  effect of the type put forward here could completely  stop  its rotation in cosmological times.
         
 \subsection{Black hole  rotational energy  diffusion}
 
  Black holes are normally expected to be   born  with  rather high  values of   spin \cite{Heger:1999ax}.  In fact  the most natural   scenario calls for  black holes to  be  formed  at  or near  extremality, simply because the   single  most important effect countering the gravitational collapse of  gravitationaly bound  matter  (for which  usual  physical pressures are  not longer   sufficiently  relevant) is the   conservation of angular momentum and the ``centrifugal'  barrier it   offers  against  gravitational collapse. Thus  the   gravitational   collapse  can   be  expected to   take  place   rather rapidly once the angular momentum barrier  is  finally overcome,  and  this  will  typically  be associated with situations where  the angular momentum  is close to maximal  for that condition. In other words  it seems  rather   natural  to assume that   a  substantial   proportion  (and  even the   great  majority) of  black holes  are  born  at  or close to extremality\footnote{Although effective mechanisms  of angular momentum transport within the BH progenitor might offer  paths  to  alter  such  generic scenario\cite{Fuller:2019sxi}.}. 
  
   On the  other hand  an extremal black  hole  does not seem to represent,  from the entropic  point of view,   a most   natural   configuration.   That is,   for a given mass,  an extremal   black hole  is  very  far  away  from  a  maximum of  entropy. In  fact,   for a given mass, the maximal  entropy would correspond to  a non-rotating  Schwarzschild  black hole.
   
    It  is thus natural to expect that  nature would  `find' some physical  mechanism   which  would  tend to  drive     situations   involving     rotating  black holes  towards  non-rotating ones.   The Penrose process   and  Hawking radiation  are two  known   effects  that  would  contribute to this. Penrose's process  however requires the  accretion of matter   with  appropriate distribution of angular momentum   from the    BH surroundings, and thus  would not be   active generically.   Hawking radiation involving preferential emission  of   quanta   carrying away angular momentum would be  another process  contributing to that,   but  it would  be  an extremely slow  one.   On the other hand a friction force,   as the one  we  are considering,  would  clearly be a process  acting in the appropriate direction  and,  as we will see,   could potentially  be a rather efficient one.  
    
From  the  point of view of the relational space-time  granularity ideas that underlays our 
approach to these issues it seems  quite natural to expect such friction-like effect. Indeed if continuity is replaced by something discrete at the Planck scale then it is 
natural to expect that (as in concrete approaches to quantum gravity) deviations from 
exact rotational invariance would open an channel for non-conservation of angular 
momentum. This would be a genuine quantum gravitational effect that would drive black holes to their maximum entropy non-rotating state. 
In other  words,   as  the   granular   aspects of   the gravitational  environment of  the  rotating  black hole  will not   poses the   exact axial  symmetry expected from the  purely classical perspective, it is natural to expect  an  associated   violation   of  conservation of   angular momentum   would ensue.  
  
   Now  we consider   what is  the amount  of energy  associated   with  the  rotation of a black hole that can become  available  for dissipation via  such process.   The answer  to this   question  is far  from  evident  given that  our effect is supposed to  have  a quantum gravity origin, and  we  of course  do not have anything like a  workable full  quantum   gravity theory to  investigate the   issue (or to derive  our hypothetical  effect).  There is however a rather  simple estimate  that can be obtained  by  considering that  during the  spin down  process  the entropy of the black hole  itself  should  not   decrease \footnote{Note however that this is not
    evident in the present context   because we   are dealing with effects of a fundamentally quantum gravitational  nature, and  as is well known   even QFT effects  might lead to violations of   the  $  dA/dt >0$ classical expectations.}.
  For concreteness we will assume from now on that the process is `adiabatic', i.e., $A=$constant. 
   That is  we will consider for simplicity the  case where all   energy   associated with  rotation that  was free to dissipate  quasi-statically  under the  constrain that  the   area  of the BH  did not  decrease.  
  
   In that case   the analysis is  simple.  The area of the event horizon  for a Kerr black hole  with mass $M$ and angular momentum  $J$  (and  thus  $ a =J/M$  ) is  given by $A(M, a)= 8\pi M(M+ \sqrt{M^2-a^2})$. 
 This means that the mass  as   function of the area is  just given by \be M= \frac{A}{8\pi}   \frac{1} {\sqrt{ \frac{A}{4\pi} - a^2}}.\ee
From the previous formula it is easy to see that in the transition from an extremal ($a^2=A/(8\pi)$) to a non-rotating black hole ($a=0$) of the same area $A$ the amount of energy made available  for dissipation is  \be\label{eff} \Delta M =\frac {2- \sqrt{2}}{2} \sqrt{ \frac A{8\pi}} \approx 0.3 \sqrt{\frac A{8\pi}} = 0.3\, M_{\rm ext},\ee where $M_{\rm ext}$ is the initial mass of the (close to) extremal black hole \footnote{This  is   a well known result. See  for instance\cite{Wald:1984rg}.}. In other words  there  would  be  up to   30\%  of the   energy density in black holes    available for dissipation under the   above discussed  conditions.  As  we see,  it is indeed the case that  (in  ordinary conditions) the translational   energy available in the peculiar motions  of  BH  in   galaxies is   simply  negligible  compared with  that available in  their rotation.   
   
 It should be noted that although the  above estimated  efficiency  bound seems  rather solid   when   considering a single   black hole     subject to the friction  torque  we are proposing, this might  not be the    total  efficiency of each single   black hole  in    its contribution to energy dissipation on  cosmological  time scales. The point is that   it  is entirely possible  that  after  two such  black holes have    lost their individual spins, they might collide    with a    high  value   of  total   angular  momentum, leading,    after the  emission of gravity waves (now possibly modified  by the  kind of  effect  we have in question),   to  a    standard  highly rotating  black hole. Such   black hole would  then  undergo the same process as     described  above,     contributing further to the   energy dissipation.  Thus   the  net  combined  efficiency  of   our process   could  be higher than the  $30\%$  estimated above.  The    detailed  analysis of this possibility   depends   again on the   evolution  and   distribution of  black holes in   galaxies   and  their  resulting collision  probabilities.

 \subsection{The explicit  form of  the  black  hole spin friction}
  We have  argued  generically that  it is natural to expect  the kind  of   friction inducing  relational granular structure of space-time  to  affect the  spin of the  rotating black hole. 
Moreover as noted   in  the discussion   around  equation \eqref{modimodi} the consideration  of a deviation from geodesic  motion such as  that  in  equation \eqref{modimodi2} cannot be done  without, at the same time including a modification of the evolution law for the spin   of the  object  in question.   

However, in contrast with the case of a fundamental particle for which the spin is a basic quantum that cannot be changed, the black hole spin is a macroscopic quantity that can be lowered via the friction mechanism with the fundamental granularity. This allows for a new term in the spin evolution equation.
Concretely,  in the case of  black holes   we   must the postulate: 
   \be\label{spinny}
    u^{\mu}\nabla_{\mu} s^{\nu}=\overline\alpha_{\rm bh} \frac{M}{m^2_{\rm p}}\, {\rm sign}(s\cdot \xi)\tilde { \mathbf R}\, (s\cdot s)\, u^{\nu}- \overline\beta_{\rm bh} \frac{M}{m^2_{\rm p}}\, \tilde { \mathbf R}_{\rm BH}\, s^{\nu},\ee
where $\overline\beta_{\rm bh}$ is a new dimensionless coupling controlling the  ``intrinsic" spin diffusion term. 
The quantity $\tilde { \mathbf R}_{\rm BH}$ is a measure of the local curvature around the black hole.  Such  quantity could be  associated to some  averaged value of the local Kretschman  scalar $\sqrt{ R_{abcd}R^{abcd}} $,  in the  region  occupied by the black hole, or something  along those lines. We will not try to be very precise  on that  at this point because it is really not necessary. Whatever the 
correct quantity would be, it will have to be of the order of the inverse square of the characteristic size of the black hole which can be defined in terms of the horizon radius\footnote{The reader   would  understandably be concerned  by the fact that there  are two 
distinct measures of curvature entering the  above  formula  referring to a single  black hole.  The point is that  the first  term  is   
intimately connected  with the one occurring in   equation (\ref{modimodi2})  which  characterizes the  translational  friction experimented  by the black hole as a result  of its translational motion  with  respect   to the matter that   generates  the   mean curvature of the  space-time. The black hole is,  of course, translationally at  rest  with respect  to  its  ``own    reference  frame" (i.e. the one  defined  by  its   own    four momentum  as  seen  from a sufficiently large distance).   Thus,  the curvature entering in that  term    should not  
involve that  associated  with the black hole itself.   On the  other hand  it is   well known  that   the rotational motion    generically is  connected  with  differential effects such   as `` dragging of the frames" and   thus  from  our point of  view,  it can  be tied to  frictional type effects between the   ``grains of space-time " that     are,    correspondingly, associated   with different   rotational  velocities. All these considerations   are of course  highly speculative  and tied to the heuristic picture outlined  here and  elsewhere\cite{Corichi:2005fw, Bonder:2007bj, Perez:2018wlo, 
Perez:2017krv}. The formulas  must therefore  be  looked upon   only as `` educated guesses" to  be applied  at the phenomenological  level. A  fundamental theory of quantum gravity  sufficiently   developed   to  work out  all the  issues in detail,   would  be a prerequisite   for a more solid   justification of these  expressions.}. 
Thus we take 
\be\label{rere}
\tilde{ \mathbf R}_{\rm BH}\approx \frac{1}{r_{\rm bh}^2}
\ee
The new term is responsible for an exponential decay of the black hole spin due to `rest frame torque'  on the  rotating black hole. Namely, the previous equation implies
 \be \label{21}u^{\mu}\nabla_{\mu} (s\cdot s)=-2\overline\beta_{\rm bh}\frac{M}{m^2_{\rm p}}\, \tilde{ \mathbf R}_{\rm BH}\, (s\cdot s),\ee
 In  other   words the  evolution of the   black hole's  spin   in  its rest frame     is  simply  $ s^2(t) =  s^2(0)  e^{- 2 t/\tau_s}$   or $ ||s(t)|| =  ||s(0) || e^{-t/\tau_s}$   with $\tau_s^{-1}  = \overline\beta_{\rm bh}\frac{M}{m^2_{\rm p}} { \mathbf R}_{\rm BH} $\footnote{In this analysis we are ignoring the time dependence of the mass implied by the effect as it would be a correction that does not affect the order of magnitude estimate.}. 
Then the  {\it life-time} of the   black  hole's spin can be estimated to be 
\be
\tau_s =
\frac{2}{\overline\beta_{\rm bh}}\frac{M}{M_{\odot}} 10^{-5} s
\ee
That is for $\overline\beta_{\rm bh}$ order unity the lifetime would be extremely small and just given by the classical BH characteristic time scale. This  could  have  rather large effects potentially observable in  LIGO which is a  rather exciting possibility\footnote{ It should  be pointed out however that  the  strict analysis of   the situation is a bit more delicate than what    was presented here. The point is that  there is a competing  process  which can  be characterized as a type of precession  driving the  spin  orientation  into  the plane that is  perpendicular to both $u$ and $\xi$. In  fact the  equation above can be  used to  compute the proper time evolution of  $ q \equiv \xi\cdot s$    which takes the  form  
$$ \frac{ dq}{d\tau}=  -B  {\rm sign }(q) - Cq + H s^0 u^0,$$ where  $  B= - \overline \alpha_{\rm bh} \frac{M}{m^2_{\rm p}}u\cdot\xi s\cdot s$,  $C=\overline\alpha_{\rm bh} \frac{M}{m^2_{\rm p}}$  and $ H $ is the Hubble parameter. We note that  as  $s^0 u^0$  has  opposite   sign  as $ q$   and   vanishes  when   $ q $ vanishes,    all the terms   tend to  decrease the value of  $ q$. Thus at  the same  time that the spin decreases   it also tends to align along   directions  orthogonal to $\xi$ ( and $u$).   However  all of the above  ignores,   of course,  possible  interactions   with other  celestial bodies  and thus a phenomenologically   precise study of the situation  will  be  much  more involved that   what we intend to  do  in this first  analysis. }.  On the other hand   life-times comparable with cosmological scales would  clearly  be   possible  if  the  coupling constant that is sufficiently weak. For instance if we take 
$\overline\beta_{\rm bh}= \sqrt{m_{\rm p}/M}=\sqrt{M_\odot/M}\,10^{-19}$ we  would  get:
  \be\label{tita}
\tau_s=2\sqrt{\frac{M}{M_{\odot}}} 10^{19-5} s\approx \frac{1}{158} \sqrt{\frac{M}{M_{\odot}}} {\rm byrs}.
\ee
Note that for this simple estimate we have used $r_{\rm BH}=2M$  which is incorrect if the black hole has spin. However, this gives the right order of magnitude and simplifies the analysis.  With the choice \eqref{rere} there will be indeed an additional spin dependence in equation (\ref{21}), which would have a more complicated solutions but with similar qualitative behavior so that the previous life-time estimate  would still be reasonable.  The previous expression is just an estimate from the natural ansatz $\overline\beta_{\rm bh}\approx \sqrt{m_{\rm p}/M}$. The correct estimate of such effect necessitates a deeper understanding of the fundamental nature of black holes in quantum gravity. The key point here is that for a small enough coupling (say $\beta_{\rm bh} \lessapprox 10^{-19} \sqrt{{M}/{M_{\odot}}}$) the numbers become interesting: the previous ansatz for $\overline\beta_{\rm bh}$ gives a spin lifetime of the order of 6 million years for a solar mass black hole while the life-time for the spin of a supermassive black hole remains of the order of and larger that the age of the universe.  But all this has to be discussed together with the effects 	on the translational degrees of freedom controlled by the magnitude of  $\overline\alpha_{\rm bh}$.

\section{Possible phenomenological consequences in galaxies}\label{peno}

Before studying the cosmological effects of our proposal on the evolution of dark energy , we will briefly discuss possible (more local) consequences regarding  galaxies. In order to organize the discussion we will only consider the two models introduced in (\ref{GW1}) and \eqref{GW2}. It is clear that, from a phenomenological perspective other possibilities should be explored. To give an idea of the type of effects that the new physics we  have put forward can have, we will concentrate on these two natural examples.

\subsection{The weaker translational coupling \eqref{GW1}}

A possible consequence of this scenario is that black holes of masses of the order of solar masses should be closer than expected to galactic centers or they may  even be falling into the central super-massive black hole. The reason is that the friction might be more important than the Newtonian force when they are spinning (recall estimate \eqref{gala}). More precisely, let us assume in this section that $\overline\alpha_{\rm bh}=\overline\alpha^{01}_{\rm bh} {m_{\rm p}}/M$ and that $\overline\beta_{\rm bh}\approx \overline\beta^0 \sqrt{m_{\rm p}/M}$ (and also take $n=0$ in \eqref{gala}). With these numbers  a solar mass BH would stop in 6 myrs, the force ratio (\ref{gala}) would be about $0.1$ at its maximum.  This is probably an effect that is too weak and   lasts too short of a time to have important consequences. However, for a 100 solar mass BH the spin will last for about 60 myrs,  and the force (\ref{gala})  would reach a maximum of about $100$ times the Newtonian binding force. Before the spin goes down and the friction disappears the energy loss via our mechanism would induce such black holes to fall towards  an orbit closer to the galactic center or even to merge  with the central black hole. On the other hand,  super-massive BHs are thought to spin for times comparable to the age of  the universe. In that case our model indicates  that those must be always in the center of galaxies. Never alone (without an important accretion disk selecting their rest frame) orbiting around other isolated sources. As far  as we  know this is  compatible with observations. 

Therefore, if the order of magnitude of couplings proposed in the previous paragraph are correct, the spin friction would be quite important for (highly spinning) black holes of the order of a few solar masses so that they would stop rotating quickly in galactic times (a few $10^2$ myrs for a the sun galactic orbit). However, the deviations from Newtonian  translational motion would not be an important effect in those cases, because the force is small and lasts for too short of a time to produce appreciable deviations.  This mechanism would imply that intermediate mass black holes found in galaxies would strongly populate the low spin part of their phase space, either because they were born with low spin or because they have lost it via the friction that we postulate here.  This  seems a rather  attractive  scenario,  phenomenologically speaking,  because of the  data suggesting that progenitor  BH's observed via gravitational wave astronomy have unexpectedly  low spins (see \cite{LIGOScientific:2018jsj}). 									

On the other hand larger mass  black holes ($M\gtrapprox 100\,  M_{\odot}$) have spins lasting longer while the translational force becomes more important according to  (\ref{gala}). Such black holes would have gone through a dramatic  slow down in  their motion around  the galaxy and, as a result  they  might   have  fallen   down  into the galactic center coalescing with  other black holes there. Such process could be relevant for the formation and/or the growth of super-massive black holes in the galactic centers. This  might account for  the  origin of  super-massive black holes,  as  well as for the   generic  correlation between the  galaxy 's mass   and the mass of the 
central  BH\cite{osti_21367367}.

A related effect would be a higher than expected  rate  of mergers between super-massive BHs in galactic centers and medium mass BHs. The gravitational signal of such events would be in low frequency range and hence would have remained undetected via gravitational wave detectors so far. However, they would produce secondary effects due to the disruption of the local accretion disks around super-massive BHs.  A  detailed  study is  required to analyze  the   expected  frequency range and establish  their possible  detectability  and expected  event rate  by,  say,   advanced  LIGO and LISA, as a function of the  model's parameter.   

One place in   which   the  effect might become relevant  is  in  BH  mergers.
The estimate $\eqref{GW1}$ implies (at least in the range of  values of couplings explored in this section) 
that such effects could become comparable to the gravitational forces for spinning black holes with $M\gtrapprox 100 M_{\odot}$, and would already be at the $3\%$ level for $M=30 M_{\odot}$. Note that these numbers would be suppressed by factors of the order of $a/M$, so high spin values is what would maximize signals. Such effects could become measurable in the near future if spinning black hole collisions are detected. 

Finally, recall that in the estimates (\ref{gala}) as well as (\ref{GW}) we are assuming the existence of a background scalar curvature $\mathbf R$ of the order of the present value of the cosmological constant, i.e., $\mathbf R\approx 10^{-120} m_{\rm p}^2$. This is important because the translational part of the friction force depends on having produced the cosmological constant by the time black holes form. This would be in agreement with the scenario of \cite{Perez:2018wlo, Perez:2017krv} as the cosmological constant is quickly generated  during the early evolution of the universe around electroweak transition time. We will see in the next section that the energy lost via the friction effects postulated in the previous section can make this initial cosmological constant evolve into a higher value, and that that can alleviate (and possibly resolve) the so-called $H_0$ tension.  These specific   phenomenological aspects   
 will be considered  in some detail in  a  companion paper\cite{Ed}. 

\subsection{The stronger translational coupling \eqref{GW2}}


 For  the  case  where the  suppression associated   with   the parameter $\overline\alpha_{\rm bh} $  in   not as  strong as  in the  previously considered  case the  effects  of our  model would  be rather  dramatic. However there will  be   strong interplay  between the  rotational   and  translational   aspects  of the  model.   That  is,   either  the  rotational friction  would  be  rather effective   and    would    bring    down   all rotation of  the  black    holes   rapidly to    essentially  vanishing values (and in particular the  progenitors   black holes  involved in  LIGO  would have  $a/M<10^{-13}$)   or    the translational friction  would    rapidly stop the rotation of the black holes   around the  galaxy,   and they  would  therefore quickly  fall  towards  the center,  merging   rapidly  with the super-massive black  hole of the  corresponding galaxy.  This  could  in  fact  be a mechanism for the  formation and  growth of such  black holes,  which  as was already  noted,   is  a question that continues to baffle  astronomers.  The detailed study of  such   scenario  clearly deserves   close  attention,   but that  task  is  beyond the scope of the present  paper.   The point, however, is that the  model is clearly  susceptible  to empirical   examination, a rather welcome  feature in  a field  where most models  seem  to be either  very hard  or   sometimes even  impossible to be explored
 observationally, beyond the  specific  feature   they   were designed to account for.

 \section{Possible resolution of the $H_0$-tension via diffusion from black holes.}
 
  The  considerations  above illustrate how the basic  equations  we proposed in \cite{Perez:2018wlo, Perez:2017krv} to describe the noisy -energy-diffusion of  fundamental particles due to Planckian granularity could  be  extended (and suitably modified) to  describe, at the effective level, a similar friction-like  effect  on  both the translational and rotational motions of black holes.  In the case of fundamental particles, and as   analyzed in detail in \cite{Perez:2018wlo, 
Perez:2017krv}, the diffusion of energy can  generate a dark energy component of the order of the present cosmological constant during the   cosmological electro-weak transition   epoch.  Similarly, if the effects we  have  put forward here apply to black holes, then the associated energy diffusion  might have a  significant  impact in the evolution of dark energy at late cosmological times\footnote{One  might, at first sight,  worry that  in  order to  deal  with  the   effect we have been considering in the context  of  black holes a complex  and  detailed study of the effect on   each individual  black hole  must be   undertaken. The point is that,  just as  we  can  proceed to  do  cosmology  at late times   considering  all  galaxies,  the  intergalactic  gas, etc.,  as  collectively representing a fluid  
  that,  at a  certain  level of  approximation  can  be   taken as  homogeneous and isotropic,   as far as  cosmological studies  are concerned, we  should be   able to proceed  in the  same  way   in dealing with the    black hole component of such fluid  and   the   dominant aspects  that our effect  would  have at  that level  of analysis.  It is  clear, of course, that the   more detailed  study  of these question at the local level,  could be expected to become much more complex,  just as the internal dynamics  of galaxies   is  a much more  complex  problem  than  that  of   estimating   their  aggregate and averaged   contribution   to the  cosmological  mean  density.  We expect to   undertake such  studies of the effects of our hypothesis on the detailed    dynamics of  black holes  in the near future.}.
However, in order to  make  a detailed analysis of the effect we would  need to  have a rather clear  and detailed understanding of the  cosmological  history of  black hole  formation,  their relative abundances, the fraction of  dark matter that   might  have been   involved in  their formation, their distribution inside  galaxies, and their  dynamical behavior during galaxy  formation and  growth.  The point is that although  enormous amount of work has gone  into modelling all these issues,  the fact is  that there seems to be  at   the present  time  no clearly unified consensual view regarding such questions. As  far  as  we know,  it is  possible,  although  it is usually  taken  as  very  unlikely,  that black holes  might have formed  directly from  dark matter clustering  in  very early times. In  such  rather speculative scenario,  the  usually expected   fragmentation of  some  gigantic  dark matter clouds  did  not take place to a  sufficient  extent, and  the gravitational collapse  of   large   enough parts of the   primordial cloud  could have led directly to the formation of  some  super-massive black holes, even  before galaxies  formed and  stars  first lightened up. The point  is,  of course, that  we   still do not have a solid  account of how  such  supermassive  black holes formed.  The  above  scenario   would   have  supermassive   black holes  forming very early   after  decoupling,  and,   for rotational  life times like the one contemplated in  equation  (\ref{tita} ),  the  energy  flux into the   dark  energy sector  could  continue  until  the present times\footnote{This  would  provide  a physical mechanism  for  the  effect  described in  equation  (\ref{cce1})  below  and  some of the models,  specufically those  involving a   continuous   energy  transfer,  which   are  described  in   more  detail  the  companion paper\cite{Ed}}. 
On the other hand   such  scenario   might   or  might  not coexist  peacefully  with  recent  surveys  indicating  rather  large    spin   values  for most super-masive black holes \cite{Reynolds:2011sba}.

 The   observational  constrains   on primordial  black   hole abundances  in fact   allow  for  over  10\% of   $\Omega_{\rm DM}$ (depending  on the precise  value of  BH mass)   to be in  the form of   black holes   for   BH  masses up to   1  solar mass  
 \cite{Katz:2018zrn}) \footnote{The  detailed  structure of the bounds   as  a function of   black hole masses can be   seen in figure 6 of \cite{Katz:2018zrn}.  We note however that   the  analysis   carried  out there  seems  to  focuss   on  a  purelly monocromatic  case ( i.e.  where all  black holes   had  the   same mass)  rather on what seems   more natural, namely a   broad  spectrum  of  black hole  mass  distribution.},   with sharp  decreases for more massive   black holes,  coming from  the CMB and for   much less massive ones,  from    micro-lensing  studies\cite{Katz:2018zrn}, as well as   bounds  on  even  smaller primordial BH's from consideration of   Hawking  radiation in the  $\gamma$-ray  part of the spectrum.   On the  other   hand there  are  intriguing  and yet unexplained  correlations  between the  galactic  structure and dynamics  and  the  central  super-massive black hole  usually  found  at the  galactic  center\cite{osti_21367367}.   All  these questions  would  need  to  be   considered    with   great care   in a   detailed  analysis of  the  precise    interplay between the effect we are proposing  and the cosmological  evolution of the  black  hole   distribution functions.

 What  we seem to  have   a  much   more   solid  handle on  is  the  amount of matter that  is  available for  black hole formation  starting at the CMB   and  the  amount  of matter that  must be left  at  late times   as indicated  by  galaxy  and galactic  cluster  mass  estimates.   The data  at this time allows for a gap.  The idea  is that a fraction of the  matter  both baryonic  and   dark   that was present at the CMB  went  to form black holes,  and that  a good portion of  that energy took  the form of  rotational and kinetic   energy  of those black holes. According to out ideas, 
  a  fraction of that  energy  was dissipated as  a results    of the friction-like    forces  and torques    
  effectively  generated by the   space-time granularity of quantum gravitational origin. In that case,  such  energy  dissipation  would  lead trough the  dynamics of uni-modular gravity to a change in the value of  the  
  cosmological constant, in conjunction with an anomalous decrease of the cosmological matter  
   density  component. That is,  the  ``cosmological constant''  would  change  in tandem with an anomalous deviation from the dust $1/a^3$   behaviour of the  matter  density. Both effects would mimic a deviation from the expected equation of state making $ w_{DE} \equiv (P/\rho)_{DE} $ (DE stands for Dark Energy) 
    seem less than $-1$,  while making $w_{\rm matter}$  appear as  anomalous  (it  would  in fact become   negative  in the  limit  where  we take matter as ``powder"  with   its ordinary pressure  strictly  vanishing). 
  
The correct interpretation comes from the modification of the continuity equation for the mater density  component  in the context of unimodular gravity,  which takes  the form:
 \be\label{cce1}
 \dot \rho_{\rm matter}+3 \rho_{\rm matter} \frac{\dot a}{a}=-\dot \rho_{\Lambda}
 \ee
 and can be written in the more convenient form
 \be\label{cce}
 \frac{d(a^3 \rho_{\rm matter})}{dt}=-\dot \rho_{\Lambda} a^3,
 \ee
 where $\rho_{\rm matter}$ and $\rho_{\Lambda}$ denote the (dark matter plus baryon) matter density and the (time dependent) dark energy energy respectively.  
 These equations follow directly from the Bianchi  identity  and equation (\ref{Lambda-Eff})  specialized to a  perfect  fluid  and  the  FLRW   cosmological case.
 
  The exact analysis  of the  evolution of   both   $\rho_{\rm matter}$  and  $\rho_{\Lambda}$  requires detailed   modeling   of  the fraction   
  \be 
  f_{\rm BH}  (t, M, a) =\frac{\rho_{\rm BH}(t,M,a)}{\rho_{\rm matter}(t)}
   \ee 
  of matter that is  present  in the  form of  black holes at each  time $t$  during the  cosmological evolution as a function of their mass $M$,  angular momentum $a$, as well  as  the  efficiency  $\epsilon (a)$ of energy  dissipation.  Taking the  simplifying assumption that   that all  black holes are born as extremal ones, that $\epsilon ( a=M) = 0.3$, and that  the process of   energy  dissipation is rather fast  on the relevant cosmological scales,   one  would   be able   to  compute things explicitly.
  
  However,   in  the  absence of a clear understanding  of the cosmological  evolution of black holes distribution,  and of a clear answer to the question  of how much  dark matter can be converted into black holes, is it still possible to make an estimate of the contribution of our mechanism to the late cosmological evolution of dark energy? It seems that, despite the uncertainties, one can still produce an encouraging and rather safe estimate in terms of two basic ingredients: On the one hand, by the requirement that a sufficient amount of matter present at the CMB has not become black holes and has remained to  account for what we see  at present  in galaxies,  galactic  clusters,  inter-stellar gas,  etc.  On the other hand, the  demand that at most $30\% $ of  all the matter that went into forming  black holes (recall  the  efficiency estimate  in equation  \eqref{eff})  would  be dissipated into feeding the dark energy component according to equation \eqref{cce}.  We  should note that it is  entirely conceivable that  an important  fraction of the  dark matter known to be there   in galaxies and  galaxy clusters takes the form of black holes (most of which, by our arguments, would be non-spinning), and that a significant fraction of that was not  directly attributable  to   baryonic  mater  which is known to represent  about  $ 0.04/ 0.3 =0.13$  of  the  total    matter   density  at the time  of  last scattering,    which   itself represented about   $75\% $  of the critical density at that time  (the dark energy contribution at that time  is negligible,  but the photons and neutrinos, unlike what ocurs  in  the  present, do   contribute  significantly).  In fact   it seems  that   the  baryonic  matter  we can  directly observe   as  luminous components (or light absorbing gas and dust)   does not   really account for all baryonic  matter that was  present at the CMB and  at nucleosyntesis \cite{Shull_2012}\footnote{Although  recent   studies indicate a  significant portion of  the missing baryons  might have been found \cite{Nicastro:2018eam}.}  and thus  it seems  attractive to  consider a scenario in which  an important fraction   went   to  form of  black holes,  with part   of that  energy  dissipated  via  the  kind  of   process  contemplated in this  work.      
  
In  a companion  paper \cite{Ed}  we have considered some simple  models of  the situation discussed above, and have found  that   such  simple scenarios    can naturally account for the   now   famous ``$ H_0$  tension". The   basic  outcome of   the  analysis  presented indicate  that   having   as little as a   few \%  of the   matter  (including  dark and baryonic  components)  originally  appearing  at  the  CMB  being dissipated  via  the process we have  outlined,  taking place  during  the  intervening period  between the  Last Scattering  Surface  (LSS) and today,  it is  possible, though the use of the   equations of uni-modular gravity  to  account for   the so  called $ H_0$ tension without  negatively  affecting  the overall  viability of the   modified  $\Lambda$CDM model  as   far as late time observations is  concerned.  Needless is to say   the  modified  model   presented  here does not imply any deviation   from  the standard   $\Lambda$CDM model regarding the conditions at the LSS or before.  
  
  \section{Discussion}
  
  We  have  considered  a  scheme that would  unify the  successful   estimate  of the   order of magnitude  of the  value of the cosmological constant   presented in previous  works with a related process having  several potential observable consequences, and which, in addition, has the potential of accounting for certain behavior of the dynamics of the late universe that  seem to indicate real tensions in the  standard $\Lambda$CDM  model.  
  
  The original  model  for a  generation  of  a  cosmological constant    relied  on the hypothesis of  a  `friction inducing'  granularity of  space-time   resulting from   quantum gravity, together with the     use of   unimodular gravity, a  theory, that, in  principle, allows  for  departures from the exact conservation of energy momentum tensor.    In the present   paper  we   extend these ideas  arguing that   while    they   would  be unsuitable for  application to   ordinary macroscopic  bodies  they   might  apply    with only  small modifications   to   black holes.  The outcome of   such     effects  would    be   the   change  during the   relative late cosmological times of the value of the cosmological   constant together  with a change in the  matter content of the   universe.
Additional consequences  of   the proposal include:  an  account for the  fact that   gravity wave detection  of black  hole colisions   up to this time seem to   involve   progenitor   black holes  with   very  low  or no  spin.  A natural account  for the   wehereabouts of the   invisible  baryonic matter\cite{Shull_2012}.  We have  mentioned   a  scenario   in which   supermassive black holes  would  form   very  early on   after  decoupling  and  loose their rotational  energy in   cosmological times  scales.  That would  offer  a  mechanism  for a continuous  change of  the    dark energy  component,  and  thus would fit   with  some  of  the detailed phenomenological  models   for   resolution of the  $H_0$ tension  described in  the comanion paper\cite{Ed}.  We  have   also   briefly considered  scenarios  in which the flux  of  energy   from the  matter to the dark  matter  sector  would start  latter on  in the  cosmological history,   once   black holes have  formed  by the usual mechanisms  (i.e.  those involving   gravitational collapse of  stars of   various  sizes  at the  end of their   normal cycle).  Such  scenarios  would   underly the  physics  behind other  models   for   resolution of the  ``$H_0$ tension"  treated  in \cite{Ed}  as well as a  natural  path  for the growth of  the super-massive black holes at the center of  galaxies, and other potential consequences discussed in Section \ref{peno}.  
  
   {Much  more  work  will be needed  to   help establishing  whether  the present account  is  supported  by   further  analysis and observations. } In the  meanwhile  we consider  the proposal as one a rather   conservative one  in  comparison  with those  currently available,  in the sense of connecting the account  for  the  rough value of the cosmological constant obtained  in   previous   works \cite{Perez:2018wlo, 
Perez:2017krv} with the details of its  late time  variation, into a  relatively unified  model,   while  requiring no  exotic equation state  for dark energy sector.  The fact that  the latter is often  postulated  in  a completely {\it add  hoc} manner with the only purpose of fitting  the data,   while offering no  clear path for  the  independent  study of the proposals,  is compounded,  in our view,  with the fact, which we see as  rather  problematical (particularly  when associated  with a classical regime),  that such  equations  of   state,   usually involve  violation of the  dominant energy 
condition\footnote{That  condition is  essentially  the statement that   energy flows  are  causal   and  its  violation  in any type of energetic  component   of the universe  would  imply  faster than  light  energy  transport. }. 
   
   We  end by acknowledging the highly speculative nature of the present proposal,  while  emphasizing one  of  is  very attractive features: The possibility of   its  empirical exploration. We do  hope   that the present  manuscript  would  stimulate other  colleagues  into contributing  to the exploration of these   and  related ideas.  
              
  
\section{acknowledgements}   
 We  thank  Marco Taoso for providing useful comments  and references, Susana Landau and Edward Wilson-Ewing for reading the manuscript and offering   helpful suggestions. DS  is grateful for the support provided by  the grant FQXI-MGA-1920 from the Foundational Questions Institute and the Fetzer Franklin Fund, a donor advised by the Silicon Valley Community Foundation.

\bibliography{referencias}

\end{document}